Robust Collimation Control of Laser-Generated Ion Beam

S. Kawata[1], M. Takano[1], D. Kamiyama[1], T. Nagashima[1], D. Barada[1], Y. J. Gu[2, 3], X. Li[2], Q. Yu[2], Q. Kong[2], and P. X. Wang[2]

[1]Utsunomiya University, 7-1-2 Yohtoh, 321-8585 Utsunomiya, Japan
[2]Fudan University, Shanghai, China
[3]ELI-Beamlines, Institute of Physics, Prague, Czech Republic

The robustness of a structured collimation device is discussed for an intense-laser-produced ion beam. In this paper the ion beam collimation is realized by the solid structured collimation device, which produces the transverse electric field; the electric field contributes to reduce the ion beam transverse velocity and collimate the ion beam. Our 2.5 dimensional particle-in cell simulations demonstrate that the collimation device is rather robust against the changes in the laser parameters and the collimation target sizes. The intense short-pulse lasers are now available, and are used to generate an ion beam. The issues in the laser ion acceleration include an ion beam collimation, ion energy spectrum control, ion production efficiency, ion energy control, ion beam bunching, etc. The laser-produced ion beam tends to expand in the transverse and longitudinal directions during the ion beam propagation. The ion beam collimation is focused in this paper.

# I. INTRODUCTION

By chirped pulse amplification, a higher laser intensity has been developed, and intense short pulse lasers are now available for various applications. On the other hand, ion beams are useful for medical ion cancer therapy, basic particle physics, controlled nuclear fusion, high-energy sources, material treatment, and so on. The energy of ions, which are accelerated in an interaction between an intense laser pulse and a target, reaches a few tens of MeV or more[1-6]. The issues in the laser ion acceleration include an ion beam collimation, ion energy spectrum control, ion production efficiency, ion energy control, ion beam bunching, etc. Depending on ion beam applications, the ion particle energy and the ion energy spectrum should be controlled as well as the ion beam quality.

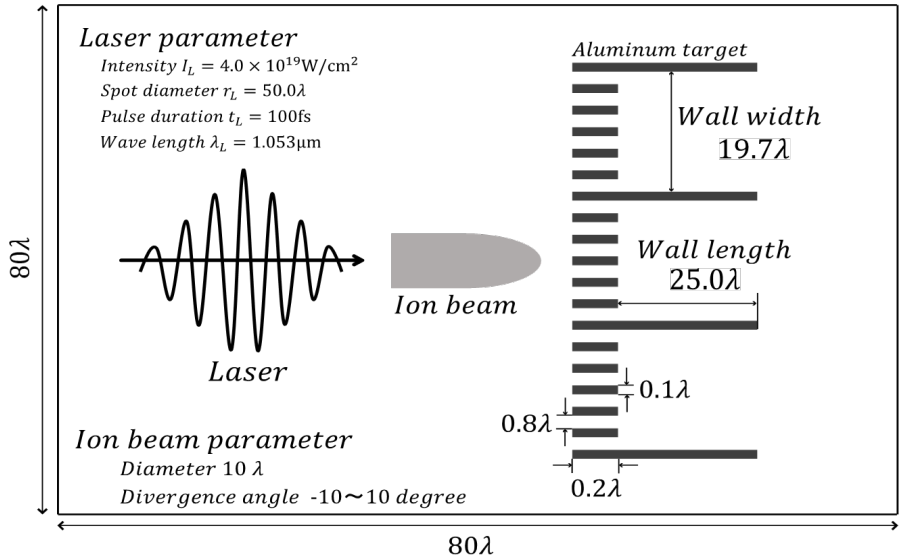

FIG. 1. Structured thin-foil collimation device for ion beam. A pre-accelerated proton beam is introduced to the collimation device. The holes' walls create the transverse electric TNSA field to collimate the ion beam.

In this paper the ion beam collimation is studied[7-8] in the laser plasma interaction, in which a thin-foil structured target is employed; the structured thin foil target (see Fig. 1) has holes behind the target to generate the transverse electric field, which contributes to the reduction of the ion beam transverse velocity toward the ion beam collimation. The structured target, which has the holes behind the target, was proposed in Refs. 7 and 8 to produce the collimated ion beam at the ion source[6]. However, the laser-produced ion beam tends to expand in the transverse direction, as well as in the longitudinal direction, during the ion beam propagation after its generation at the ion source. In this paper the structured thin-foil target with the holes at the target rear side[7-8] is employed to collimate the laser pre-produced ion beam. The 2.5-dimensional particle-in-cell simulations are performed to investigate the ion beam collimation.

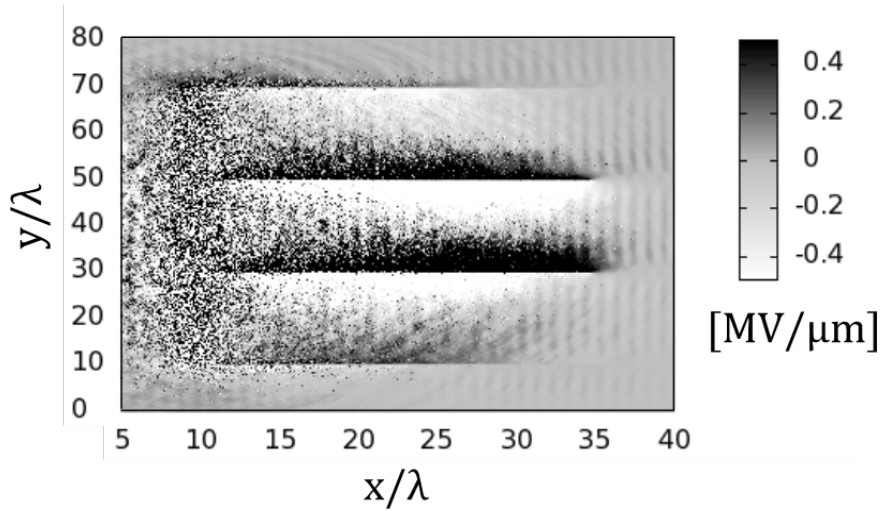

FIG. 2. The transverse electric field. The Al structured target is illuminated by an intense laser. The fine structure absorbs the laser energy efficiently, and generates high-energy electrons. The TNSA transverse field is generated by the electrons collimates the proton beam.

## II. ROBUST COLLIMATION OF ION BEAM IN LASER TARGET INTERACTION

Figure 1 shows the collimation device and the simulation setup employed in this paper. The sub-wavelength fine structure at the thin foil itself increases the laser absorption ratio[9-11], and the larger holes at the target rear creates the transverse TNSA (target normal sheath acceleration) field to collimate the ion beam. The ion beam diameter of $10\lambda$ is smaller than the rear hole diameter of $19.7\lambda$ (the wall width in Fig. 1) to avoid the ion beam split[7-8]. The pre-accelerated proton beam is shown in Figs. 3(a) and (d), and is introduced to the simulation box initially. The pre-accelerated proton beam has the divergence angle shown in Fig. 4(a) and the energy and spatial distributions in Figs. 3(a) and (d). The Gaussian laser intensity is $4.0 \times 10^{19} W/cm^2$, the laser spot size is $50\lambda$ and the laser pulse length is 100fs. Figure 2 shows the transverse collimation electric field at the target during the laser illumination. Figures 3(a)-(c) show the ion beam distributions at $t$=0fs, 300fs and 580fs without the collimation device, respectively. Figures 3(d)-(f) present a typical result for the collimation at $t$=0fs, 300fs and 580fs, respectively. Figures 3(d)-(f) demonstrate that the collimation device is effective for the ion beam collimation. Figure 4 shows the ion beam divergence angle, and presents the clear collimation of the ion beam by the structured target.

We have investigated robustness of the collimation device against the changes in the laser intensity, the laser illumination timing, the structured target wall length and the target wall width. The definitions for the target wall length and width are presented in Fig. 1. Figures 5(a)-(d) summarize the results for the parameter studies to find the robustness of the collimation device.

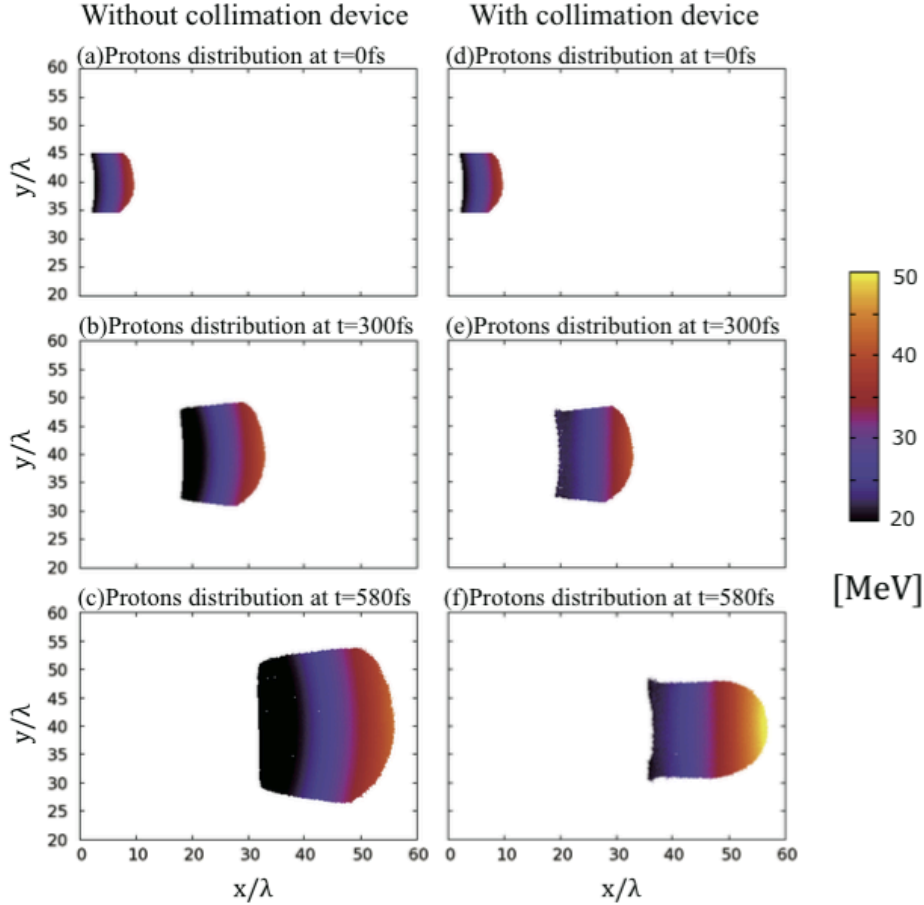

FIG. 3. Spatial distributions of protons for the proton beam without the collimation device at (a)$t$=0fs, (b)$t$=300fs and (c)$t$=580fs, and for the proton beam with the collimation device at (d)$t$=0fs, (e)$t$=300fs and (f)$t$=580fs. The collimation device reduces the proton beam divergence successfully.

Figure 5(a) shows the relation between the ion beam divergence angle and the laser intensity. When the laser intensity is weak compared with the optimal laser intensity, the ion beam collimation field is weak. On the other hand when the laser intensity is too high, the ion beam is over-focused. Figures 6 present the beam proton distributions and the proton divergence angle distributions for the laser intensity shown in the figure. The allowance of the laser intensity is rather wide as presented in Fig. 5(a). Figure 5(b) shows the relation between the ion beam divergence angle and the target wall width (see Fig. 1). For a specific laser intensity, when the wall width becomes narrow, the transverse collimation electric field becomes strong and so the collimation effect becomes too strong. When the wall width is too large, the electric field felt by the beam ions becomes small. Figures 7 shows typical example simulation results for the different target wall widths. In this paper the ion beam diameter is $10\lambda$. Figures 5(b) and 7 show that the wall width should be the order of the beam diameter. Figure 5(c) shows the relation between the ion beam divergence angle and the target wall length (see Fig. 1). When the target wall length is too short, the time period, during which the beam protons feel the collimation transverse electric field, becomes short. So the collimation effect becomes weak for the short target wall length. When the target wall length becomes too long, a part of the long wall near the laser target interaction area in which the target has a fine structure has the collimation electric field. So the rest of the long target wall has no effect for the collimation, and the too-long wall does not degrade the collimation effect. Figures 8 present the typical simulation results in this case. Figure 5(d) presents the relation between the ion beam divergence angle and the laser timing. When the laser timing is too fast, just the head of the ion beam is collimated. When the laser timing is late, the tail of the ion beam is collimated. Figures 9 show the typical simulation results for the corresponding laser timing. The laser pulse length is 100fs in this paper, and the tolerable laser-timing shift is rather wide as shown in Fig. 5(d).

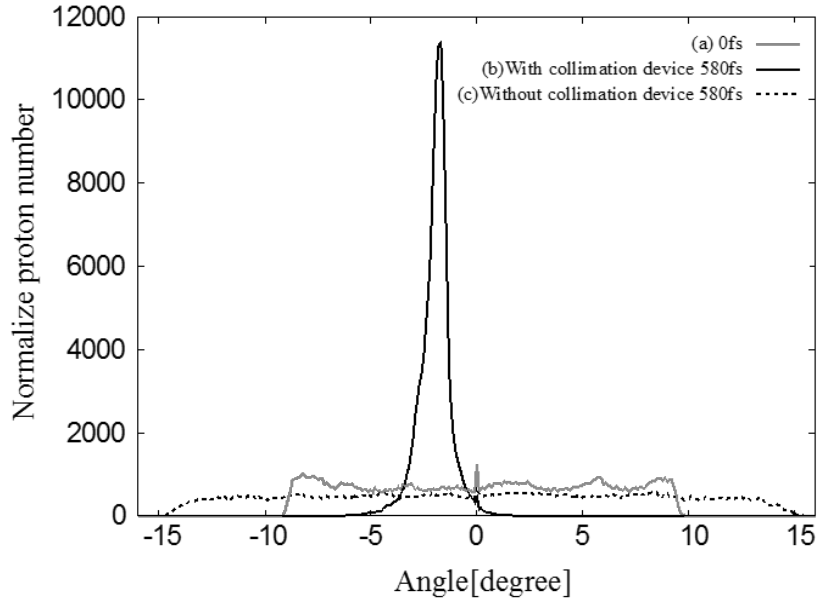

FIG. 4. Divergence angle distributions for (a) the original proton beam (solid line), for (b) the proton beam without the collimation device (short-dotted line) and for (c) the proton beam with the collimation device (long-dotted line).

All the parameter study results are summarized in Figs. 5 and demonstrate that the collimation device of the structured target is rather robust against the changes in the laser parameter and the target parameter. The results presented in this paper show the viability of the laser-based collimation device for the collimation of the pre-accelerated ion beam, which may be produced by an intense-laser target interaction.

III. CONCLUSIONS

We have presented the robustness of the structured collimation target for the pre-accelerated ion beam against the laser and target parameter changes. In the collimation device the transverse electric field is employed to collimate the diverging beam ions and is generated by the intense-laser target interaction. The collimation device would be required for a long distance transportation of the ion beam pre-accelerated[6]. For actual applications of the ion beam, the ion beam should be transported in a long distance compared with the size of the laser target interaction area. The collimation device proposed contributes the real ion beam use and would be viable for a realistic long-distance ion beam transportation. For a practical laser-based ion beam accelerator[6], the multiple acceleration, collimation and bunching stages would be required to meet the application requirements for the ion beam quality, energy and intensity. The collimation device shown in this paper is one of the promising candidates for the pre-accelerated ion beam collimation.

234801.

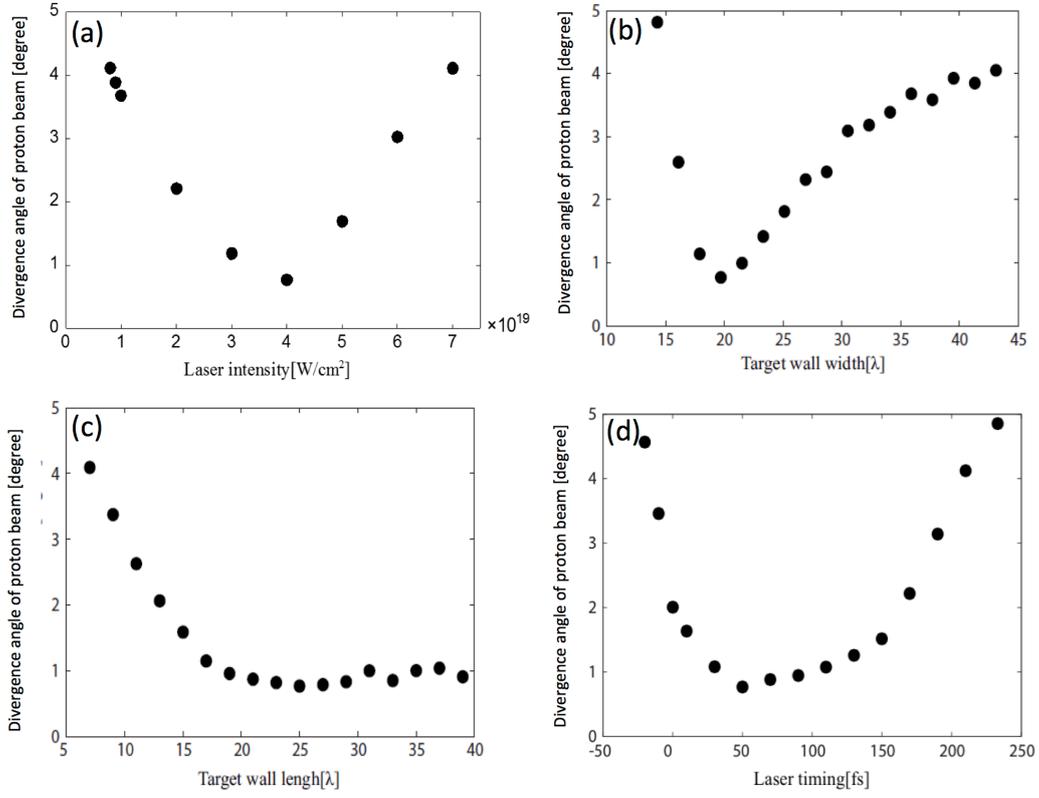

FIG. 5. Divergence angle dependences of beam ions on (a) the laser intensity, (b) the target wall width, (c) the target wall length and (d) the laser timing. The collimation device shown in Fig. 1 is rather robust against the parameter changes in the laser parameters and the target structure.

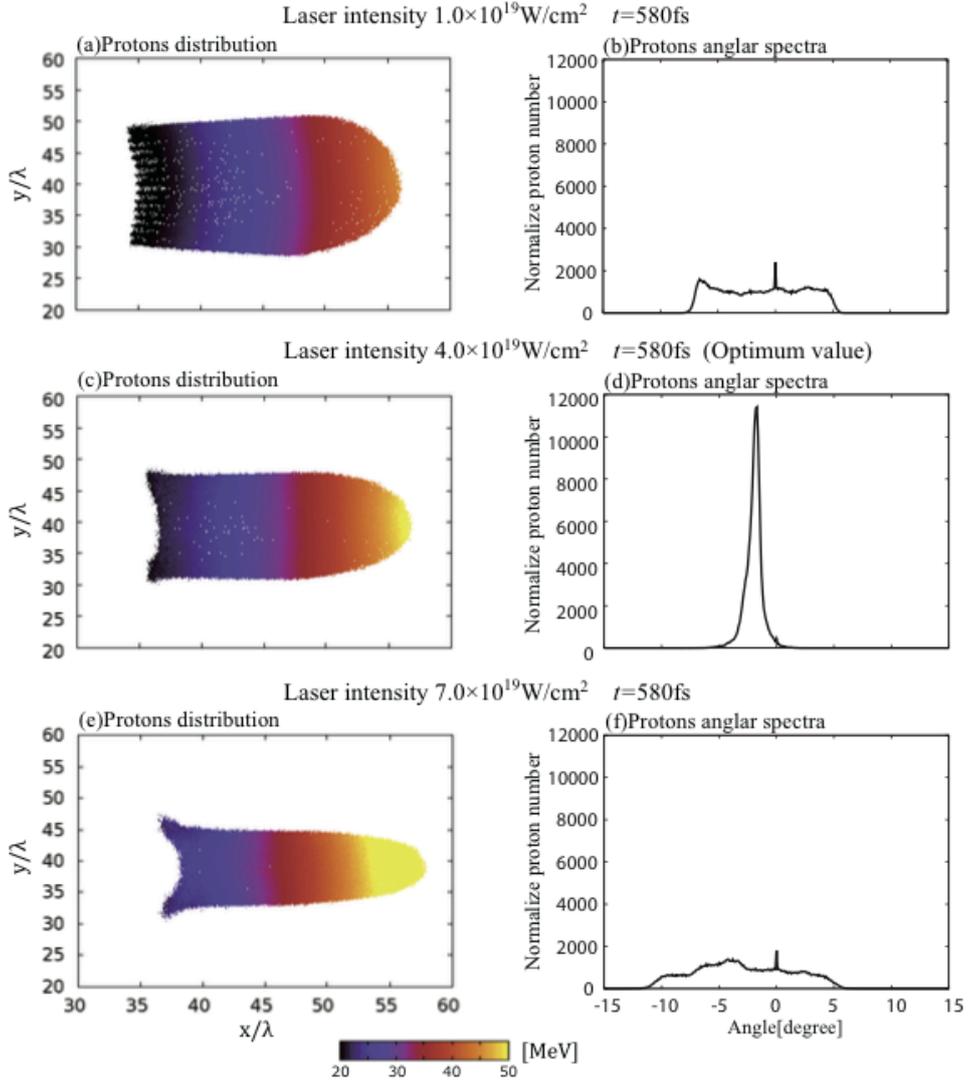

FIG. 6 Distributions of the beam protons at the laser intensity of (a) $1.0\times10^{19}W/cm^2$, (b) $4.0\times10^{19}W/cm^2$ and (c) $7.0\times10^{19}W/cm^2$ at $t$=580fs. The corresponding distributions of the proton divergence angle are shown in (d), (e) and (f) at $t$=580fs.

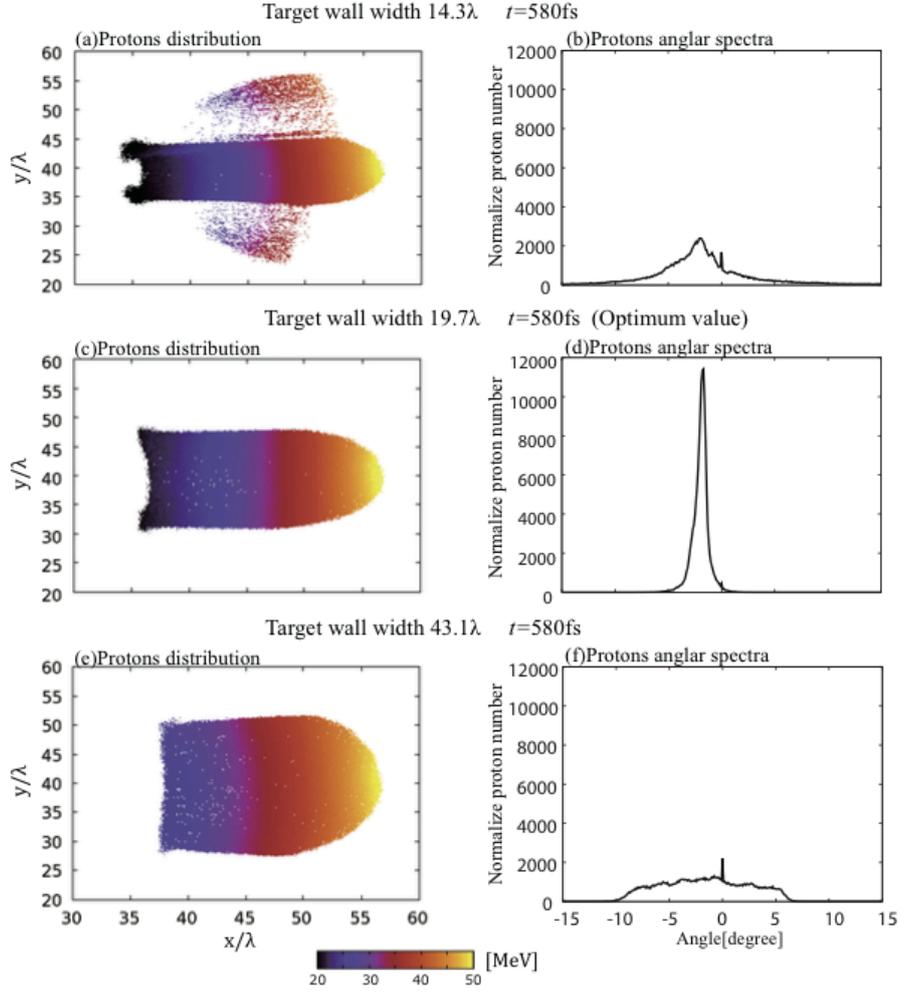

FIG 7. Distributions of the beam protons at the target wall width of (a) 14.3$\lambda$, (b) 19.7$\lambda$ and (c) 43.1$\lambda$ at $t$=580fs. The corresponding distributions of the proton divergence angle are shown in (d), (e) and (f) at $t$=580fs.

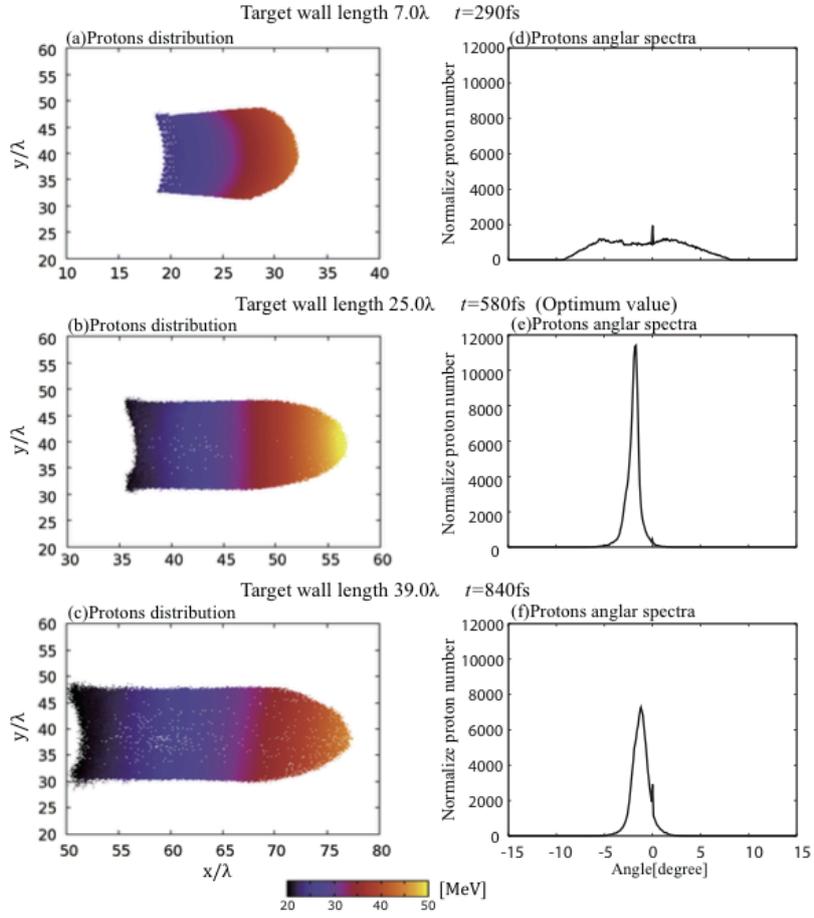

FIG. 8 Distributions of the beam protons at the target wall length of (a) 7.0$\lambda$ at $t$=290fs, (b) 25.0$\lambda$ at $t$=580fs and (c) 39.0$\lambda$ at $t$=840fs. The corresponding distributions of the proton divergence angle are shown in (d), (e) and (f).

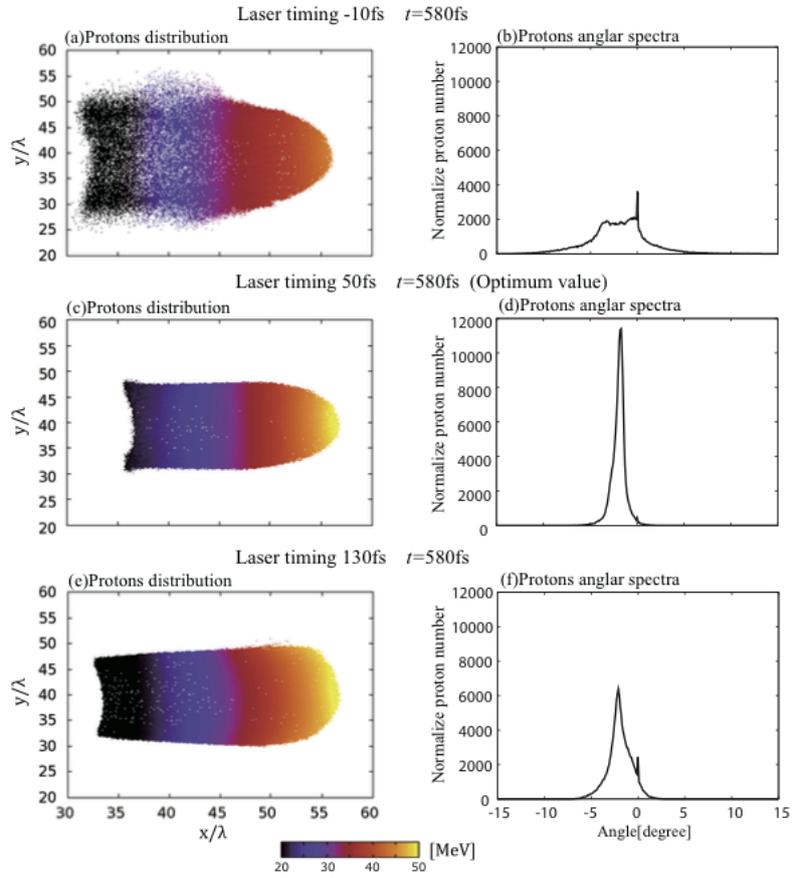

FIG. 9 Distributions of the beam protons at the laser timing of (a) -10fs, (b) 50fs and (c) 130$fs$ at $t$=580fs. The corresponding distributions of the proton divergence angle are shown in (d), (e) and (f) at $t$=580fs.